\documentclass[preprint,showpacs,preprintnumbers,amsmath,amssymb]{revtex4}
\usepackage{amsmath}

\usepackage{graphicx}% Include figure files
\usepackage{pdfpages}
\usepackage{dcolumn}% Align table columns on decimal point
\usepackage{bm}% bold math
\usepackage{amsmath} 

\newcommand{\qed}{\nobreak \ifvmode \relax \else
      \ifdim\lastskip<1.5em \hskip-\lastskip
      \hskip1.5em plus0em minus0.5em \fi \nobreak
      \vrule height0.75em width0.5em depth0.25em\fi}

\begin{document}

\preprint{}

\title{Octonionic two-qubit separability probability conjectures}
\author{Paul B. Slater}
 \email{slater@kitp.ucsb.edu}
\affiliation{%
Kavli Institute for Theoretical Physics, University of California, Santa Barbara, CA 93106-4030\\
}
\date{\today}% It is always \today, today,
            
\begin{abstract}
We study, further, a conjectured formula for generalized two-qubit Hilbert-Schmidt separability probabilities that has recently been proven by Lovas and Andai (https://arxiv.org/pdf/1610.01410.pdf) for its  
real (two-rebit) 
asserted value ($\frac{29}{64}$), and that has also been very strongly supported numerically for its complex ($\frac{8}{33}$), and quaternionic ($\frac{26}{323}$) counterparts. Now, we seek to test the presumptive {\it octonionic} value of $\frac{44482}{4091349} \approx 0.0108722$. We are somewhat encouraged by certain numerical computations, indicating that this (51-dimensional) instance of the conjecture might be fulfilled by setting a certain determinantal-power parameter $a$, introduced by Forrester (https://arxiv.org/pdf/1610.08081.pdf), to 0 (or 
possibly near to 0). Hilbert-Schmidt measure being the case $k=0$ of random induced measure, for $k=1$, the corresponding octonionic separability probability conjecture is $\frac{7612846}{293213345} \approx 0.0259635$, while for $k=2$, it is $\frac{4893392}{95041567} \approx 0.0514869, \ldots$.
The relation between the parameters $a$ and $k$ is explored.
\end{abstract}


\maketitle

\tableofcontents
\section{Introduction}
For the (Dyson-index-like \cite{MatrixModels}) values $\alpha = \frac{\beta}{2} =\frac{1}{2},1, 2$, corresponding to  real, complex and quaternionic scenarios, this (``concise'') formula (\cite[eqs. (1)-(3)]{slaterJModPhys}),
\begin{equation} \label{Hou1}
P_1(\alpha) =\Sigma_{i=0}^\infty f(\alpha+i),
\end{equation}
where
\begin{equation} \label{Hou2}
f(\alpha) = P_1(\alpha)-P_1(\alpha +1) = \frac{ q(\alpha) 2^{-4 \alpha -6} \Gamma{(3 \alpha +\frac{5}{2})} \Gamma{(5 \alpha +2})}{3 \Gamma{(\alpha +1)} \Gamma{(2 \alpha +3)} 
\Gamma{(5 \alpha +\frac{13}{2})}},
\end{equation}
and
\begin{equation} \label{Hou3}
q(\alpha) = 185000 \alpha ^5+779750 \alpha ^4+1289125 \alpha ^3+1042015 
\alpha ^2+410694 \alpha +63000 = 
\end{equation}
\begin{equation}
\alpha  \bigg(5 \alpha  \Big(25 \alpha  \big(2 \alpha  (740 \alpha
   +3119)+10313\big)+208403\Big)+410694\bigg)+63000,
\end{equation}
as well as this second, rather differently-appearing formula \cite[p. 26]{SlaterFormulas} (with 
$k$ set to zero),
\begin{align*} \label{Hyper1}
P_2\left(\alpha\right)   &  =1-\frac{\alpha\left(  20\alpha+8k+11\right)
\Gamma\left(  5\alpha+2k+2\right)  \Gamma\left(  3\alpha+k+\frac{3}{2}\right)
\Gamma\left(  2\alpha+k+\frac{3}{2}\right)  }{2\sqrt{\pi}\Gamma\left(
5\alpha+2k+\frac{7}{2}\right)  \Gamma\left(  \alpha+k+2\right)  \Gamma\left(
4\alpha+k+2\right)  }\\
&  \times~_{6}F_{5}\left(
%TCIMACRO{\QATOP{1,\frac{5}{2}\alpha+k+1,\frac{5}{2}\alpha+k+\frac{3}{2},2\alpha+k+\frac{3}%
%{2},3\alpha+k+\frac{3}{2},\frac{5}{2}\alpha+k+\frac{19}{8}}{\alpha+k+2,4\alpha+k+2,\frac{5}%
%{2}\alpha+k+\frac{7}{4},\frac{5}{2}\alpha+k+\frac{9}{4},\frac{5}{2}\alpha+k+\frac{11}{8}}}%
%BeginExpansion
\genfrac{}{}{0pt}{}{1,\frac{5}{2}\alpha+k+1,\frac{5}{2}\alpha+k+\frac{3}{2}%
,2\alpha+k+\frac{3}{2},3\alpha+k+\frac{3}{2},\frac{5}{2}\alpha+k+\frac{19}{8}%
}{\alpha+k+2,4\alpha+k+2,\frac{5}{2}\alpha+k+\frac{7}{4},\frac{5}{2}\alpha+k+\frac{9}{4},\frac
{5}{2}\alpha+k+\frac{11}{8}}%
%EndExpansion
;1\right)
\end{align*}
{\it both} yield to arbitrarily high-precision: that 
$P_1(\frac{1}{2})=P_2(\frac{1}{2})=\frac{29}{64}$; that $P_1(1)=P_2(1)=\frac{8}{33}$; and that $P_1(2)=P_2(2)=\frac{26}{323}$.
(The variable $k$ in the formula for $P_2(\alpha)$, which we set to zero for 
our [Hilbert-Schmidt-based] discussion here, parameterizes a broader class of (``random induced'' \cite{Induced}) measures--which itself will be a focus of some discussion later in the paper.)

Lovas and Andai  \cite{lovasandai} have very recently proven that the first ($P_1(\frac{1}{2})=P_2(\frac{1}{2})=\frac{29}{64}$) of these three cases, in fact, corresponds to the probability that a random (with respect to Hilbert-Schmidt/Euclidean/flat measure \cite{szHS,ingemarkarol}) pair of real quantum bits (``rebits'' \cite{carl}) is separable/unentangled. They also gave an integral formula, they hoped would prove the second ($\frac{8}{33}$) case for pairs of complex (standard) quantum bits. Taking a highly-intensive numerical approach, Fei and Joynt \cite{FeiJoynt2} have found supporting evidence for the three (real, complex and quaternionic) cases.

Uninvestigated, so far, however, has been the case $\alpha=4$, for which $P_1(4)=P_2(4)= \frac{44482}{4091349} \approx 0.0108722$ (with, $44482 = 2 \cdot 23 \cdot 967$ and $4091349 = 3 \cdot 29 \cdot 31 \cdot 37 \cdot 41$). This, motivated by random matrix theory \cite{2016arXiv161008081F}, with $\alpha =\frac{\beta}{2}$, with $\beta$ being the usual ``Dyson-index'',  would appear to possibly correspond to an octonionic setting 
\cite{2016arXiv161008081F}.

So, our objective in this paper is to find/construct  a framework in which to address 
the conjecture that $P_1(4)=P_2(4)= \frac{44482}{4091349}$ has a valid octonionic interpretation.

Let us note that the pairs of rebits constitute a 9-dimensional space of $4 \times 4$ density matrices--nonnegative definite, symmetric with real entries and unit trace. The pairs of complex (standard) quantum bits similarly constitute a 15-dimensional space, and the pairs of quaternionic quantum bits, a 27-dimensional space. The pairs of octonionic quantum bits would comprise a 
51-dimensional space.

The two formulas $P_1(\alpha)$ and $P_2(\alpha)$ were developed based solely on analyses of matrices with 
real and complex (and not quaternionic and octonionic) entries. To be more detailed, the ascending 
moments of determinants of the (real and complex) 4 x 4 density matrices and of the determinants of their 
``partial transposes'' were computed (first, for Hilbert-Schmidt [$k=0$] measure), and formulas found for them. These were, then, used in the 
Mathematica density approximation procedure of Provost \cite{Provost}, 
to eventually arrive at the expressions for $P_1(\alpha)$ and $P_2(\alpha)$. (Typically, well more than
the first 
ten thousand moments were employed to arrive at high-precision estimates of rational-valued separability
probabilities. Then, the FindSequenceFunction command of Mathematica was utilized with series of these values 
in helping in the process of constructing
the underlying formulas $P_1(\alpha)$ and $P_2(\alpha)$.)

The two original (real and complex) moment formulas  
(Charles Dunkl observed \cite[App. D]{MomentBased}) could be absorbed into one, 
by regarding the parameter ($\alpha$) in the complex case to be twice that in the real case 
(hence, the apparent [Dyson-index-like] connection to random matrix theory).

Now, although the calculation of determinants is certainly a well-developed  subject with matrices the entries of which are restricted to real and complex values, it becomes more subtle with the (non-commutative) quaternions, and, a fortiori, it would seem with matrices composed of the (non-commutative and non-associative) octonions. 
E. H. Moore \cite{moore1922determinant} gave a definition in the 
quaternionic  case, while the ``Dieudonne determinant \ldots is a generalization of the determinant of a matrix over division rings and local rings'' \cite{ wiki:xxxx}. Also, the concept of ``quasideterminant''  
(work of Israel Gelfand {\it et al} \cite{gelfand2003quaternionic}) appears relevant in this regard. 
In a series of extensive 2012 unpublished analyses of Dunkl,  the appropriateness of the Moore determinant in this quaternionic context found strong support.
(Let us note that Fei and Joynt  appear to have 
by-passed the use of determinants, in their quaternionic analysis \cite{FeiJoynt2}.)
Further, S. Alesker asserted ``for octonionic hermitian matrices of size at least 4, no nice notion of determinant is known, 
while for matrices of size 3 it does exist'' \cite{alesker} (but cf. \cite{LiaoWangLi}). 

So, the issue at hand is whether or not the  moment formulas Dunkl developed \cite[App. D]{MomentBased} 
can be validly ``extrapolated''and applied meaningfully to the octonionic domain. 
\section{Analyses}
The question at hand pertains to $4 \times 4$ (density) matrices ($\rho$)--and, in this regard, we seek to extend the quite recent analyses of $2 \times 2$ and $3 \times 3$ ``Wishart matrices ($W$) with octonion entries'' of Peter Forrester \cite[sec. 3]{2016arXiv161008081F}. (The trace-normalization condition for
density matrices will not be of concern here.) He employed Cholesky decompositions $W=T^{\dagger} T$ (cf. \cite{HouChai}). Accordingly, we start with  $4 \times 4$ null matrices $T$ and fill, their six upper triangular off-diagonal entries with octonions, the eight independent components of each of the six being distributed as standard Gaussians. 

Next, we fill the four diagonal entries with values that are the square roots of gamma distribution variates. For the $2 \times 2$ case, Forrester employed $\Gamma[a+1,2]$ and $\Gamma[a+5,2]$. In the $3 \times 3$ instance, he utilized $\Gamma[a +4 (i-1),2]$, $i=1,2,3$. (The parameter $a$--akin to the determinant power $k=K-N$ in the random induced measure formula \cite[eq. (3.6)]{Induced}--appears to not need to be specified in advance in the Forrester presentation, but must be large enough that the Gamma distributions are well-defined.) For the $4 \times 4$ case, we employed $\Gamma[a+4 (i-1),2]$, $i=1,\ldots,4$. (It was not fully clear to us if
$\Gamma[a+4 (i-1)+1,2]$, $i=1,\ldots,4$ might be a [more?]  appropriate alternative 
[cf. \cite[eq. (2)]{diaz2016riesz}]).

To further proceed, we relied upon the suite of Mathematica programs made available by Tevian Dray and Corinne A. Manogue in their paper, ``Finding octonionic eigenvectors using Mathematica" \cite{dray1998finding}. This allowed us (using their MMult command) to generate random Wishart matrices $W=T^{\dagger} T$.

For each such matrix, we sought to compute (if possible) its ``determinant'' and test its positivity (and that of the ``determinant'' of its partial transpose). For this purpose, 
we employed ``Theorem 5.3. (Laplace expansion)'' in the 2010 paper of Jianquan Liao, Jinxun Wang and Xingmin Li , entitled ``The all-associativity of octonions and its applications'' 
\cite{liao2010all}. (We followed the ``template'' of the 
Laplace expansion of a $4 \times 4$ matrix by  
$2 \times 2$ ``complementary minors'' presented in \cite{ wiki:xxx}, and utilized the 
Mathematica command Odet[X] in the Dray-Manogue package for the computation of the $2 \times 2$ minors. These minors appeared to be always real-valued in our computations.)

At this point, we were prepared for our simulation of the Wishart matrices $W=T^{\dagger}T$. At first, we set the gamma distribution parameter $a$ to 1. For 500,000 such random matrices, we found (using the  Laplace expansion algorithm) all but one of their determinants to be (numerically) considered positive. (Let us note that Forrester in simulating ten thousand $3 \times 3$ Wishart matrices, found 5,500 of them to have negative determinants--but the particular value of $a$ employed was not indicated. We also do not know what mathematical software was employed.)

Now, to address the underlying/central question of the value of the Hilbert-Schmidt separability probability of two-qubit density matrices with octonionic entries, we (again, using the Laplace expansion routine) computed the determinants of the partial transposes of the 499,999 (positive determinant) Wishart matrices. Of the 
499,999 partial transposes, 354,404 had positive determinants. Employing the two-qubit version of the Peres-Horodecki conditions \cite{peres1996separability,michal,augusiak}, this would give us a separability probability of 0.708809, orders of magnitude larger than the conjectured value of $P_1(4)=P_2(4) = \frac{44482}{4091349} \approx 0.0108722$.

But now, we ascertained that one could readily ``tune'' the separability probability by the choice of the parameter $a$. So, for $a=\frac{1}{175}$, we obtained 491,320 Wishart matrices with positive determinants, again out of 500,000 generated. Of these, only 5,127 had ``positive partial transposes'' (PPT's), giving us a separability probability of $\frac{5127}{491320} \approx 0.0104352$, 
just slightly {\it smaller} than the conjectured value. Again, with 500,000 matrices generated, using $a=\frac{1}{160}$, we obtained an estimated separability probability of 0.0114942, 
now slightly {\it larger} than the conjecture.
\section{Discussion}
We find some encouragement for these analyses, indicative of a 
(limiting?) zero (near-zero) value of the parameter $a$, from formula (1.5) 
of the cited Forrester article \cite{2016arXiv161008081F},
\begin{equation} \label{Forrester1.5}
(\lambda_1 \lambda_2)^a e^{-c (\lambda_1 +\lambda_2)} (\lambda_2 -\lambda_1)^8    
\end{equation}
 ("$c$ is simply a scale factor").
This term (\ref{Forrester1.5}) is proportional to the eigenvalue PDF (probability distribution function) for the $N=2$ case of Wishart octonionic matrices. By setting $a=0$, the term $(\lambda_1 \lambda_2)^a$--the $a$-th power of the determinant--reduces to unity. The remaining (``eigenvalue-repulsion/Vandermonde'')
factor $(\lambda_2 -\lambda_1)^8$ then remains present--while the determinant itself is removed--just as in 
the {\it Hilbert-Schmidt}-type formula \cite[eqs. (3.11), (7.5)]{szHS}, we are attempting to implement in  octonionic form.

Now, we observe that in \cite[App. F]{SlaterFormulas}, a series of formulas $P(k,\alpha)$ was computed ($\alpha=1,\ldots,75$)
giving the two-qubit separability probabilities, where the $\alpha$'s, as above, appears to correspond to
$\frac{\beta}{2}$, where $\beta$ is the conventional Dyson index. Further, $k=K-4$, where the dimension of the space in which the $4 \times 4$ density matrices are viewed as embedded is $4 K$ \cite{Induced}. So, it seems that the determinantal-power parameters $a$ and $k$ may function as transforms of one another. 
The ``octonionic''($\alpha=4$) formula reported in \cite[App. F]{SlaterFormulas} takes the form
\begin{equation} \label{Ratio}
P(k,4)=1-\frac{2^{2 k+25} (k+11) (k+12) (k+13) S \Gamma \left(k+\frac{27}{2}\right) \Gamma (2
   k+23)}{315 \sqrt{\pi } \Gamma (3 k+42)},
\end{equation}
where
\begin{equation}
S=8 k^{10}+736 k^9+30908 k^8+785888 k^7+13511051 k^6+165605534 k^5+1478827827
   k^4+
\end{equation}
\begin{equation} \label{Ratio3}
   9572954872 k^3+43203702816 k^2+122897189520 k+166878079200.
\end{equation}
We have $P(0,4)=\frac{44482}{4091349}$, the conjectured octonionic two-qubit  Hilbert-Schmidt separability 
probability. Further, $P(1,4)=\frac{7612846}{293213345} \approx 0.0259635$, a somewhat larger value,\ldots
Now, $P(12,4)=\frac{326023943703}{463672957769} \approx 0.703133$, a value close to the estimate of 
0.708809, we obtained above, when $a$ was set to 1. So, it would be an exercise of interest to
try to determine a functional relation, say $f(k)=a$, between $a$ and $k$, where $f(12) \approx 1$.

Along such lines, Proposition 4 of Forrester \cite{2016arXiv161008081F} asserts that if $W=X^{\dagger} X$, where $X$ is an $n \times 2$ matrix with random octonion entries, the eight independent components in each being distributed as standard Gaussians, then, the associated $16 \times 16$ real symmetric matrix $\omega(W)$ has two eight fold degenerate eigenvalues and their  probability distribution function (PDF) is proportional to (\ref{Forrester1.5}), as given above, with $a=4 n -5$ and $c=\frac{1}{2}$. The parameter $n$ appears to have the 
same sense as the $K$ ($=k+4$) above.

In a supplementary analysis, again with 500,000 matrices generated, now with $a=\frac{1}{2}$, all had positive determinants. Of these 252,612 had PPT's, giving us a separability probability of 0.505224. Additionally, we tested if $|\rho| >|\rho^{PT}|>0$, finding that this was the case with probability 0.817079.
Further, $P(9,4)=\frac{10180551}{20361434} \approx 0.499992$, so for our hypothesized function, relating 
$a$ and $k$,
$f(9) \approx \frac{1}{2}$. Also, for this $k=9$, the formulas given in \cite[Apps. E,F]{SlaterFormulas} yield
that the probability of $|\rho| >|\rho^{PT}|>0$, is $\frac{P(9,4)-Q(9,4)}{P(9,4)} \approx 0.932124$. 
Here (cf. (\ref{Ratio})-(\ref{Ratio3})),
\begin{equation}
Q(k,4)=
   \frac{1}{2}-\frac{2^{2 k+23} (k+11) \left(k^3+34 k^2+402 k+1608\right) \Gamma
   \left(k+\frac{19}{2}\right) \Gamma \left(k+\frac{23}{2}\right) \Gamma
   \left(k+\frac{27}{2}\right)}{\pi  \Gamma (k+17) \Gamma \left(2 k+\frac{43}{2}\right)}.
\end{equation}

As a further exercise, for each $a= \frac{1}{8},\frac{1}{4},\frac{3}{8},\ldots \frac{9}{8}$, we computed the associated
separability probability. Then, we asked Mathematica to construct a function interpolating this list.
Extrapolating the function to the value $a=0$, yielded 0.010108, while the conjectured value is 
$ \frac{44482}{4091349} \approx 0.0108722$. Similarly, extrapolating the hypothesized $f(k)=a$ function gave
$f(0) = 2.04852$.

We had observed above that 
in  the $2 \times 2$ octonionic case, Forrester \cite{2016arXiv161008081F} employed for the two diagonal entries, using the Cholesky decomposition, $\Gamma[a+1,2]$ and $\Gamma[a+5,2]$. In the $3 \times 3$ instance, he utilized 
$\Gamma[a +4 (i-1),2]$, $i=1,2,3$. However, in the $3 \times 3$ case, we speculated  that he might possibly have intended instead $\Gamma[a +1+4 (i-1),2]$, $i=1,2,3$ (cf. \cite[eq. (2)]{diaz2016riesz}). Then, in the $4 \times 4$ case of immediate interest to us, it might be more appropriate to employ $\Gamma[a +1+4 (i-1),2]$, $i=1,2,3. 4$, rather than the 
$\Gamma[a +4 (i-1),2]$, $i=1,2,3,4$ we have used in the analyses so far reported above. 
Using this modification for the diagonal entries, we computed 
(again, for 500,000 random matrices) the nine associated separability probabilities for 
$a=-\frac{99}{100},\ldots,-\frac{999}{1000}$ in intervals of $\frac{1}{1000}$. The estimate of 0.011026 closest to the  octonionic Hilbert-Schmidt separability probability conjecture was obtained at $a=-0.994$.

\begin{acknowledgments}
I am grateful to Charles Dunkl for bringing the indicated paper of P. J. Forrester to my attention. The subject of this paper was initially raised on the mathematics, physics, and Mathematica stack exchanges.

\end{acknowledgments}

\bibliography{Octonions}% Produces the bibliography via BibTeX.

\end{document}